\documentclass{article}
\usepackage{smc2025}
\usepackage[caption=false, font=footnotesize]{subfig}
\usepackage{paralist}
\usepackage[figure,table]{hypcap}

\usepackage[whole]{bxcjkjatype}

\usepackage{alphabeta}
\usepackage{arabtex}
\usepackage[LFE,LAE,LGR,T2A,T1]{fontenc}
\usepackage[greek, russian, main=english]{babel}

\usepackage{amsmath,amssymb,amsfonts}
\usepackage{algorithmic}
\usepackage{graphicx}
\usepackage{textcomp}
\usepackage{xcolor}

\usepackage[binary-units]{siunitx}
\usepackage{float}
\usepackage{tikz}
\usetikzlibrary{shapes.geometric, arrows, positioning}
\usepackage{pgfplots}
\usepgfplotslibrary{groupplots,dateplot}
\usepackage{booktabs}
\usepackage{microtype}  

\def\note[#1#2#3]{#1\if b#2$\flat_#3$\else\if#2##$\sharp_#3$\else$_#2$\fi\fi}

\def\papertitle{Beat and Downbeat Tracking in Performance MIDI Using an End-to-End Transformer Architecture}

\author[1,2]{\mbox{\firstname{Sebastian}\lastname{Murgul}\email{sebastian.murgul@klang.io}}}
\author[2]{\mbox{\firstname{Michael}\lastname{Heizmann}\email{michael.heizmann@kit.edu}}}

\affil[1]{\institution{Klangio GmbH}\city{Karlsruhe}\country{Germany}\affiliationtype{Company}}
\affil[2]{\department{Institute of Industrial Information Technology}\institution{Karlsruhe Institute of Technology}\city{Karlsruhe}\country{Germany}\affiliationtype{University}}

\completesetup

\title{\papertitle}
\begin{document}
\capstartfalse
\maketitle
\capstarttrue

\begin{abstract}
  Beat tracking in musical performance MIDI is a challenging and important task for notation-level music transcription and rhythmical analysis, yet existing methods primarily focus on audio-based approaches. This paper proposes an end-to-end transformer-based model for beat and downbeat tracking in performance MIDI, leveraging an encoder-decoder architecture for sequence-to-sequence translation of MIDI input to beat annotations. Our approach introduces novel data preprocessing techniques, including dynamic augmentation and optimized tokenization strategies, to improve accuracy and generalizability across different datasets. We conduct extensive experiments using the A-MAPS, ASAP, GuitarSet, and Leduc datasets, comparing our model against state-of-the-art hidden Markov models (HMMs) and deep learning-based beat tracking methods. The results demonstrate that our model outperforms existing symbolic music beat tracking approaches, achieving competitive F1-scores across various musical styles and instruments. Our findings highlight the potential of transformer architectures for symbolic beat tracking and suggest future integration with automatic music transcription systems for enhanced music analysis and score generation.
\end{abstract}

\section{Introduction}\label{sec:introduction}
Beat tracking aims to detect the underlying rhythmic grid within a musical performance \cite{davies2021tempobeatdownbeat}. This rhythmic grid consists of downbeats, beats, and tatum subdivisions. Downbeats refer to the first beat within a bar and therefore indicate the beginning of a new bar. In a notation-level music transcription system, we implicitly or explicitly need the rhythmical grid in order to be able to derive discrete note values from the detected tones \cite{benetos2019automatic}. In general, we have two input options in a transcription setting: We can either use the original input audio or the output performance MIDI data of a previous note tracking step.

Audio-based beat tracking has seen substantial progress, particularly with the introduction of deep learning techniques. Early approaches relied on statistical models, but modern methods increasingly leverage neural networks to improve accuracy and robustness.
One of the first notable deep learning approaches was proposed by Böck et al.\ (2011), who pioneered the use of bidirectional Long Short-Term Memory (LSTM) networks to classify beats from an audio spectrogram, smoothing predictions using autocorrelation \cite{bock2011enhanced}.
This work evolved in 2016 when Böck et al.\ proposed an RNN-based beat tracking method that outputs beat and downbeat features directly from magnitude spectrograms. A Dynamic Bayesian Network (DBN) was then used to model variable-length bars and align the detected beats \cite{bock2016joint}.
Davies et al.\ enhanced Böck's approach in 2019 by replacing LSTMs with a Temporal Convolutional Network (TCN) featuring dilated convolutions along the temporal axis \cite{davies2019temporal}. Zhao et al.\ introduced Beat Transformer in 2022, employing time-wise and instrument-wise attention mechanisms alongside dilated self-attention and demixed spectrograms to improve beat detection \cite{zhao2022beat}. Foscarin et al.\ proposed Beat This! in 2024, an advanced transformer-based beat tracker demonstrating high accuracy and generality across diverse musical styles. Since Beat This! does not rely on DBN postprocessing, it is also suitable for pieces with time-signature changes or high tempo variations \cite{foscarin2024beatthis}.

Unlike audio-based beat tracking, which has been extensively researched, MIDI-based methods have remained relatively scarce. Traditional approaches to MIDI-based beat tracking often relied on rule-based heuristics and statistical models, but recent research has begun incorporating deep learning techniques.
Cambouropoulos et al.\ proposed a system for joint beat detection and rhythm quantization in 2000 \cite{cambouropoulos2000midi}. Their approach clustered inter-onset intervals for beat detection, followed by assigning note onsets to the closest points on a metrical grid and assigning note values based on inter-onset intervals.
Temperley's 2007 book \lq Music and Probability\rq extended Bayesian probabilistic approaches to infer complete metrical grids rather than score positions relative to a bar \cite{temperley2007music}. Cogliati et al.\ presented an HMM-based system in 2016 for joint estimation of meter, harmony, and stream separation, combined with a distance-based quantization algorithm \cite{cogliati2016transcribing}.
Foscarin et al.\ introduced a parse-based system in 2019 employing weighted context-free grammars (WCFGs) for joint rhythm quantization and music score production \cite{montiel2019parse-based}. Shibata et al.\ proposed a piano transcription system in 2021 that incorporated HMMs and Markov Random Fields (MRFs) for rhythm quantization, leveraging non-local musical statistics to infer global parameters \cite{shibata2021non-local}.
Liu et al.\ proposed a Convolutional Recurrent Neural Network (CRNN)-based system in 2022 for MIDI-to-score conversion, incorporating onset-based beat detection and rhythm quantization \cite{liu2022performance}. Kim et al.\ developed a transformer and Convolutional Neural Network (CNN)-based guitar transcription model in 2023 that produced note-level transcriptions from spectrograms using beat information \cite{kim2022notelevel}. Beyer et al.\ introduced a performance MIDI-to-score conversion approach in 2024 based on the Roformer architecture. Their encoder-decoder model directly generated MusicXML tokens while implicitly performing beat estimation and rhythm quantization on MIDI token sequences \cite{beyer2024end}.

Despite these efforts, modern transformer architectures have seen limited application in symbolic beat tracking. Existing methods either rely on traditional probabilistic models or use deep learning approaches not specifically optimized for MIDI beat tracking. This gap presents an opportunity to explore more advanced techniques for symbolic music.

In this work, we introduce a novel end-to-end transformer-based approach for beat tracking in MIDI performances, achieving state-of-the-art performance and surpassing existing symbolic beat tracking methods. By leveraging modern transformer architectures, our model effectively captures temporal dependencies and outperforms previous HMM-based and neural network-based systems.
\section{Methodology}
\label{sec:methodology}

\usetikzlibrary{arrows.meta}
\usetikzlibrary{shapes.arrows}

\colorlet{blockbg}{red!5}
\colorlet{tokenbg}{black!5}

\begin{figure*}[t]
  \footnotesize
  \centering
  \vspace{-1cm} 
  \begin{tikzpicture}
    \node[draw, fill=blockbg, minimum width=3cm, minimum height=.7cm, rounded corners=.2cm] (encoder) at (1.5-0.23, .65) {Encoder};

    \node[draw, fill=gray!50, rounded corners=.05cm, above=0.25cm of encoder,  minimum width=2.8cm, minimum height=1.0cm] (encoding) {};
    \node[above=0.1cm of encoding, font=\bfseries] {Encoding};

    \foreach \x in {0,1,2,3,4,5,6,7,8} {
        \draw[fill=cyan!40] (\x *0.3, 1.5) rectangle +(0.15, 0.65);
        \draw[latex-] (0.075 + \x *0.3, 1.5) -- +(0, -0.5);
      }

    \foreach \x in {0.5, 1, 1.5, 2, 2.5} {
        \draw (3.5+\x, 1.9) edge[in=270,out=90,-latex,looseness=2,gray!60] (4+\x,2.7-3.2);
      }
    \foreach \x in {0.5, 1, 1.5, 2, 2.5, 3} {
        \draw[-latex] (3.5+\x, 1) -- ++(0, .5);
        \draw[-latex] (3.5+\x, -.2) -- ++(0,.5);
      }

    \node[draw, fill=blockbg, minimum width=3cm, minimum height=.7cm, right=1.0cm of encoder, rounded corners=.2cm] (decoder) {Decoder};

    \node[single arrow, draw=black, fill=gray!50, minimum width = 1cm, single arrow head extend=.1cm, minimum height=1cm, left=0.2cm of encoder] (input_arrow) {};
    \node[single arrow, draw=black, fill=gray!50, minimum width = 1cm, single arrow head extend=.1cm, minimum height=1cm, right=0.2cm of decoder] (output_arrow) {};

    \node[draw, fill=tokenbg, minimum width=3cm, minimum height=.7cm, align=left, left= 0.2cm of input_arrow] (input) {ON$\langle55\rangle$ T$\langle0.01\rangle$ OFF$\langle55\rangle$ \\T$\langle0.44\rangle$  ON$\langle62\rangle$ T$\langle0.44\rangle$ \dots};
    \node[above=0.2cm of input, font=\bfseries] {MIDI-like Input Tokens};

    \node[draw, fill=tokenbg, minimum width=3cm, minimum height=.7cm, align=left, right= 0.2cm of output_arrow] (output) {B$\langle1\rangle$ T$\langle0.01\rangle$  B$\langle2\rangle$ \\T$\langle0.44\rangle$ B$\langle3\rangle$ T$\langle0.89\rangle$ \dots};
    \node[above=0.2cm of output, font=\bfseries] {Beat Output Tokens};

    \draw (encoding.east) edge[out=0,in=180,-latex] (decoder.west);


    \draw[fill=gray!70] (3.425+0.5, -.5) rectangle +(0.15, 0.4);
    \foreach \x in {0.5, 1, 1.5} {
        \draw[fill=red!70] (3.425+\x, 1.5) rectangle +(0.15, 0.4);
        \draw[fill=red!70] (3.425+\x, -.5) rectangle +(0.15, 0.4);
      }
    \foreach \x in {2, 2.5, 3} {
        \draw[fill=blue!70] (3.425+\x, 1.5) rectangle +(0.15, 0.4);
        \draw[fill=blue!70] (3.425+\x, -.5) rectangle +(0.15, 0.4);
      }
    \draw[fill=gray!70] (3.425+0.5, -.5) rectangle +(0.15, 0.4);
    \draw[fill=red!70] (3.425+2, -.5) rectangle +(0.15, 0.4);
    \node[above=1.36cm of decoder, font=\bfseries] {Autoregressive Sampling};

  \end{tikzpicture}
  \vspace{-1.5cm} 

  \caption{Our model is based on the T5 encoder-decoder transformer architecture \cite{raffel2019t5} and uses MIDI-like tokens as input and outputs tokenized beat and downbeat information. During inference, autoregressive sampling with beam search is used.}
  \label{fig:model}
\end{figure*}
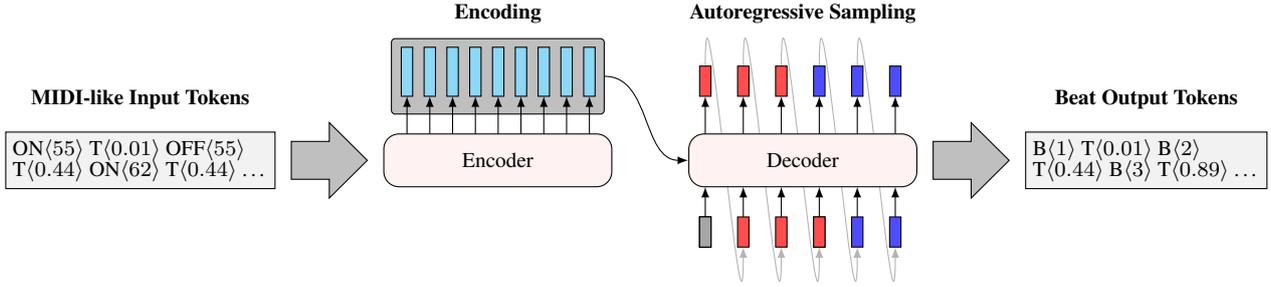

Our proposed approach for performance MIDI beat tracking is visualized in Figure \ref{fig:model}. The model follows an encoder-decoder transformer architecture, designed to translate an input MIDI segment into the corresponding beat sequence. Since transformers operate on text-based token sequences, MIDI data must first be preprocessed and tokenized before being fed into the model. The following sections detail the preprocessing pipeline, data augmentation strategies, encoding schemes, and model architecture.

\subsection{Preprocessing}
\label{sec:preprocessing}

\begin{figure}[h]
  \footnotesize
  \centering

  \begin{tikzpicture}[
      auto,
      node distance=0.5cm,
      roundnode/.append style={ellipse, draw=red!60, fill=red!5, very thick, minimum width = 2cm, minimum height = 1.2cm, text width=4.5em, align=center},
      process/.append style={rectangle, rounded corners, minimum width=2cm, minimum height=1cm, align=center, draw=black, fill=black!5, inner sep=5pt},
      arrow/.style={thick,->,>=stealth}]

    \node (midi-file)[roundnode] {A-MAPS MIDI};
    \node (extract) [process, right = of midi-file] {Extract Notes \\ and Beats};
    \node (segment) [process, right = of extract] {Split Overlapping \\ Segments };
    \node (cleaning) [process, below = of segment] {Data Cleaning};
    \node (csv-file) [roundnode, left = of cleaning] {CSV-File};
    \node (augment) [process, left = of csv-file] {Data \\ Augmentation};
    \node (quantize) [process, below = of augment] {Quantize \\ Note Times};
    \node (encoding) [process, right = of quantize] {Encoding and \\ Tokenization};
    \node (dataset) [roundnode, right = of encoding] {Dataset};

    \draw [arrow] (midi-file) -- (extract);
    \draw [arrow] (extract) -- (segment);
    \draw [arrow] (segment) -- (cleaning);
    \draw [arrow] (cleaning) -- (csv-file);
    \draw [arrow] (csv-file) -- (augment);
    \draw [arrow] (augment) -- (quantize);
    \draw [arrow] (quantize) -- (encoding);
    \draw [arrow] (encoding) -- (dataset);

  \end{tikzpicture}
  \caption{Flowchart of the data processing steps}
  \label{fig:data_processing}
\end{figure}
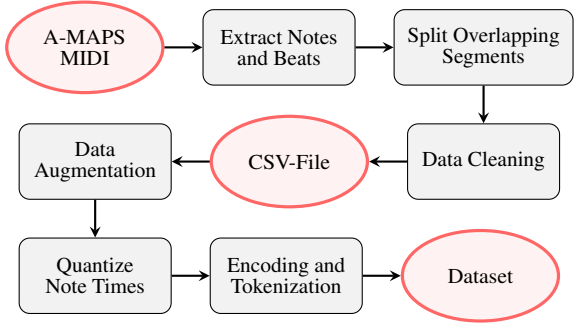

The data processing is shown in the flow chart in Figure \ref{fig:data_processing}. As input, we use the MIDI files of the A-MAPS dataset. Firstly, we extract the notes and the beat annotations from the MIDI files using the PrettyMIDI library \cite{raffel2014prettymidi}. The extracted annotations are then split into segments of $\SI{10}{\second}$ with a hop size of $\SI{1}{\second}$. In the next step, the segments are cleaned, and all examples with less than one beat are dropped. Finally, the resulting examples are stored in a CSV file for access during training.

\subsection{Data Augmentation}
\label{sec:augmentation}

The data augmentation is done dynamically in the data loading process during the training. To enrich the diversity of the examples, we perform pitch transposition and time manipulation.
The transposition augmentation shifts all the pitches of the piece within a given pitch range from \note[A0] to \note[C8]. The shift is drawn from a uniform distribution. This ensures that there is no bias for specific pitches in the training data.
One main advantage of using symbolic input over audio in a beat tracking deep learning model is the ease of manipulating the temporal properties of the data in order to increase the covered range.
For timing manipulation, we apply a random shift within the range of $\left[-\SI{1}{\second}, +\SI{1}{\second}\right]$ and a randomly drawn scaling factor between 0.9 and 1.1

\subsection{Data Encoding}
\label{sec:encodings}

In order to be able to efficiently tokenize the MIDI and beat segments for use with a transformer model, the used vocabulary should be as slim as possible. Therefore, we first quantize the absolute time values to $\SI{10}{\milli\second}$ steps. This leads to $1,000$ tokens for segments of $\SI{10}{\second}$ length.
The amount of time tokens can be further reduced by encoding relative instead of absolute time information, but our early experiments showed that this comes with the downside of error propagation and leads to overall worse results.
To evaluate the effects of data encoding on the beat tracking performance, we conducted an experiment with four different encoding strategies. Each encoding was designed to capture the essential musical information, and we experimented with different levels of abstraction and granularity. 

The input sequence of encoding \textit{v1} uses ON$\langle$\#$\rangle$ tokens to denote the pitch of a note's onset at the absolute time which is denoted by the T$\langle$\#$\rangle$ token. Similarly, the target sequence uses B$\langle$\#$\rangle$ tokens to encode the beat counter value at the beat time T$\langle$\#$\rangle$. Here, B$\langle1\rangle$ specifies the first beat, which corresponds to the downbeat. Version \textit{v2} adds a note offset token OFF$\langle$\#$\rangle$ and  \textit{v3} also a token for the note's velocity value VEL$\langle$\#$\rangle$ to the input sequence. The input encoding in version \textit{v4} is the same as in \textit{v3} but the target encoding uses only downbeat DB and beat B tokens instead of an explicit beat counter. 
Lastly, encoding \textit{v5} mirrors \textit{v2} but adopts the target representation of \textit{v4}, removing explicit beat counters while maintaining note offset encoding. This version seeks to balance structural clarity with a compact target format.
Table \ref{tab:encodings} shows short examples of the input and target sequence for each of the text encoding versions.

\begin{table*}[t]
  \centering
  \begin{tabular}{  l  l  l  }
    \toprule
    \textbf{ID} & \textbf{Input}                                                                                                                                                 & \textbf{Target}                                                                                                                   \\
    \midrule
    \textit{v1} & ON$\langle55\rangle$ T$\langle0.01\rangle$ ON$\langle62\rangle$ T$\langle0.44\rangle$ \dots                                                                    & B$\langle1\rangle$ T$\langle0.01\rangle$ B$\langle2\rangle$ T$\langle0.44\rangle$ B$\langle3\rangle$ T$\langle0.89\rangle$ \dots  \\

    \textit{v2} & ON$\langle55\rangle$ T$\langle0.01\rangle$ OFF$\langle55\rangle$ T$\langle0.44\rangle$  ON$\langle62\rangle$ T$\langle0.44\rangle$ \dots                       & B$\langle1\rangle$ T$\langle0.01\rangle$  B$\langle2\rangle$ T$\langle0.44\rangle$ B$\langle3\rangle$ T$\langle0.89\rangle$ \dots \\

    \textit{v3} & ON$\langle55\rangle$ T$\langle0.01\rangle$ VEL$\langle80\rangle$ OFF$\langle55\rangle$ T$\langle0.44\rangle$  ON$\langle62\rangle$ T$\langle0.44\rangle$ \dots & B$\langle1\rangle$ T$\langle0.01\rangle$  B$\langle2\rangle$ T$\langle0.44\rangle$ B$\langle3\rangle$ T$\langle0.89\rangle$ \dots \\

    \textit{v4} & ON$\langle55\rangle$ T$\langle0.01\rangle$ VEL$\langle80\rangle$ OFF$\langle55\rangle$ T$\langle0.44\rangle$  ON$\langle62\rangle$ T$\langle0.44\rangle$ \dots & DB T$\langle0.01\rangle$  B T$\langle0.44\rangle$ B T$\langle0.89\rangle$ \dots                                                   \\

    \textit{v5} & ON$\langle55\rangle$ T$\langle0.01\rangle$ OFF$\langle55\rangle$ T$\langle0.44\rangle$  ON$\langle62\rangle$ T$\langle0.44\rangle$ \dots                       & DB T$\langle0.01\rangle$  B T$\langle0.44\rangle$ B T$\langle0.89\rangle$ \dots                                                   \\
    \bottomrule
  \end{tabular}
  \caption{Examples of the different input MIDI-to-text and target beat-to-text encodings. ON / OFF define the pitch and T the time of the event. Additionally, the VEL token defines the velocity of the played note. The B and DB tokens define the beat and downbeat positions in the target sequence.}
  \label{tab:encodings}
\end{table*}

\subsection{Model}
\label{sec:model}

The model uses the T5 transformer architecture \cite{raffel2019t5} since the architecture already proved to be suited in other Music Information Retrieval (MIR) tasks like transcription \cite{hawthorne2021sequence}.
The Hugging Face Transformers package \cite{wolf2019huggingface} is used for implementing the network. We employ a reduced architecture of the T5 model, halving the configuration of \textit{t5-small} with a model dimension $d_\textit{model}=128$, feedforward dimension $d_\textit{ff}=1024$, three encoder-decoder layers and four attention heads.

This model is trained from scratch, utilizing the Adafactor optimizer \cite{shazeer2018adafactor} with a self-adaptive learning rate. The models are trained for $50$ epochs with a batch size of $32$. Training the model on an NVIDIA Tesla V100 takes about $6$ hours on average.

For the autoregressive sampling during inference, we found out that using a beam search with $5$ beams leads to the best results. We also specify that all ngrams of size $2$ can only occur once.

\section{Experiments}
\label{sec:experiments}

In this section, we describe the datasets used for training and evaluation, followed by the metrics used to assess model performance.

\subsection{Datasets}
\label{sec:datasets}
For the training and evaluation of the model, datasets containing synchronized MIDI and beat annotations are essential. The A-MAPS dataset \cite{ycart2018maps} is an extension of the MAPS database \cite{emiya2009maps} which consists of $270$ piano pieces with audio and corresponding MIDI annotations. The original MAPS dataset has been widely used to train and evaluate piano transcription performance. The A-MAPS dataset augments the MIDI annotations by adding rhythm (including beat and downbeat positions) as well as key annotations. Because of the significant large size and well-aligned annotations, it is used as main dataset in our studies.
The ASAP dataset \cite{foscarin2020asap} is a dataset of aligned musical scores and MIDI performances with additional beat and time signature annotations, amongst others. Therefore, it is used as a second piano dataset in Section \ref{sec:comparison}.
For a more diverse quantitative evaluation, we also look at the guitar datasets GuitarSet \cite{xi2018guitarset} and the Leduc dataset \cite{edwards2024francois}.
While GuitarSet comes with beat and downbeat annotations in the JAMS files, the orginal Leduc dataset is focused solely on the guitar transcription task. The Leduc dataset comprises 239 jazz guitar performances with accompanying high-quality transcriptions written by François Leduc in the GuitarPro\footnote{\url{https://www.guitar-pro.com}} format. The scores are converted into MIDI and aligned with the original audio using the approach described by Riley et al.\ \cite{riley2024high}. This alignment process is further refined to also adjust downbeat and beat information from the GuitarPro files according to the resulting note mapping between the score and the performance. Using this extension, we also get beat annotations alongside the aligned MIDI transcriptions.
The created beat annotations are available online and can be used alongside the original Leduc dataset files\footnote{\url{https://github.com/klangio/midi-beat-tracking}}.

\subsection{Evaluation Metrics}
\label{sec:metrics}
To quantitatively assess model performance in both studies and comparative evaluations, we rely on F1-scores for beat and downbeat detection, along with continuity metrics.
\begin{itemize}
  \item \textbf{Beat F1-score ($f_\text{b}$):} Measures the accuracy of detected beat positions relative to ground-truth annotations.
  \item \textbf{Downbeat F1-score ($f_\text{db}$):} Evaluates the model’s ability to correctly identify downbeats, which mark the beginning of musical bars.
\end{itemize}
We use the standard tolerance window of $\SI{70}{ms}$ for both beats and downbeats, as defined by the mir\_eval Python library \cite{raffel2014mireval}. Unlike some previous studies, we do not exclude the first five seconds of each sequence before evaluation, as our model operates on relatively short input sequences.
\section{Results}
\label{sec:results}

This section presents the results of our proposed method, evaluated using the experimental setup described in Section \ref{sec:experiments}. We first analyze the impact of various hyperparameters and modeling choices in an ablation study, followed by a comparative evaluation against existing state-of-the-art approaches.

\subsection{Ablation Study}
\label{sec:ablation}

In the ablation study, we show the effect of various hyperparameters, encoding schemes, and data augmentation strategies. Each model is trained for $50$ epochs on the training split of the A-MAPS dataset, while the results are obtained by evaluating on the test split of the same dataset.
If not specified otherwise, we use the T5 architecture with a segment length of $\SI{10}{\second}$, a quantization of $\SI{10}{\milli\second}$, \textit{v3} encoding, and no data augmentations.

\subsubsection{Data Encodings}
\label{sec:abl_encoding}

The choice of the text encoding is crucial for the model's ability to capture the context and rules the beat tracking estimation implicitly underlies. In Table \ref{tab:encoding_results} we show the downbeat and beat tracking performance for the encoding schemes introduced in Section \ref{sec:encodings}.
It can be observed that the model benefits from the more information it receives as input. Adding velocity as well as offset information leads to an $\SI{18}{\percent}$ higher $f_\text{b}$ score than when only relying on note onset information. This improvement aligns with the expectation that downbeats are typically emphasized, often through increased velocity values, making them more distinguishable. However, a key limitation of this approach is that velocity values are not always available, particularly for instruments like the guitar, which may reduce its applicability in certain contexts.

The best results for the downbeats are achieved with the \textit{v4} encoding. Here, we observe a great impact by the removal of the beat counter. We assume that the model already counts the beats implicitly and that having to output a specific counter value leads to confusion when the MIDI segment is cropped in a way that it does not begin with the first beat. Interestingly, the $f_\text{b}$ score drops slightly in comparison to the \textit{v3} encoding. In version \textit{v5}, on the other hand, we do not see that effect in $f_\text{db}$.

\begin{table}[h]
  \centering
  \begin{tabular}{lcc}
    \toprule
    \textbf{Encoding Scheme} & $f_\text{b}$                    & $f_\text{db}$                   \\
    \midrule
    \textit{v1}              & $\SI{81.06}{\percent}$          & $\SI{34.75}{\percent}$          \\
    \textit{v2}              & $\SI{90.11}{\percent}$          & $\SI{47.69}{\percent}$          \\
    \textit{v3}              & \textbf{$\SI{96.03}{\percent}$} & $\SI{59.52}{\percent}$          \\
    \textit{v4}              & $\SI{94.84}{\percent}$          & \textbf{$\SI{67.16}{\percent}$} \\
    \textit{v5}              & $\SI{81.23}{\percent}$          & $\SI{47.31}{\percent}$          \\
    \bottomrule
  \end{tabular}
  \caption{Comparison of beat ($f_\text{b}$) and downbeat ($f_\text{db}$) F1-scores across different encoding schemes. The results highlight the impact of incorporating velocity, offset information, and the removal of the explicit beat counter on beat tracking performance.}
  \label{tab:encoding_results}
\end{table}

\subsubsection{Segment Length}
\label{sec:abl_segment_length}

The choice of the segment length does not only have a direct impact on the sequence lengths the model has to process, but also affects the vocabulary size via the number of time tokens needed. Therefore, having a shorter segment length reduces complexity with the disadvantage of reducing the context window, too. Table \ref{tab:segment_length_results} shows the results for different segment lengths.
For the beat F1-score, the segment length of $\SI{10}{\second}$ leads to the best results. Since the $\SI{15}{\second}$ length performs better than the $\SI{5}{\second}$ length, it can be assumed that the sweet spot for the aforementioned tradeoff lies somewhere in the interval $[\SI{10}{\second},\SI{15}{\second})$.

\begin{table}[h]
  \centering
  \begin{tabular}{lcc}
    \toprule
    \textbf{Segment Length} & $f_\text{b}$                    & $f_\text{db}$                   \\
    \midrule
    $\SI{5}{\second}$       & $\SI{91.77}{\percent}$          & \textbf{$\SI{65.54}{\percent}$} \\
    $\SI{10}{\second}$      & \textbf{$\SI{96.03}{\percent}$} & $\SI{59.52}{\percent}$          \\
    $\SI{15}{\second}$      & $\SI{94.99}{\percent}$          & $\SI{57.10}{\percent}$          \\
    \bottomrule
  \end{tabular}
  \caption{Impact of segment length on beat ($f_\text{b}$) and downbeat ($f_\text{db}$) F1-scores. The results illustrate the tradeoff between shorter segments, which reduce computational complexity, and longer segments, which provide a broader temporal context for beat tracking.}
  \label{tab:segment_length_results}
\end{table}

\subsubsection{NLP Task Interpretation}
\label{sec:abl_nlp_task}

Our implementation using the T5 transformer interprets the task of beat tracking as a translation between the MIDI language and the rhythm language. Alternatively, the text completion interpretation can also be applied by using the GPT2 model \cite{radford2019gpt2}. Here, the model is trained to generate beat tokens for a given sequence of MIDI tokens in a text completion manner. Therefore, a `MIDI:' token followed by the MIDI notes and a `Beat:' token is used as a primer sequence. The GPT2 model now completes the text by adding the beat token sequence. A comparison of the results from the GPT2 model with the T5 model is shown in Table \ref{tab:nlp_task}. We can see that the T5 clearly outperforms the GPT2 model in terms of beat and downbeat F1-scores.

\begin{table}[h]
  \centering
  \begin{tabular}{lcc}
    \toprule
    \textbf{Model Architecture} & $f_\text{b}$                    & $f_\text{db}$                   \\
    \midrule
    T5                          & \textbf{$\SI{96.03}{\percent}$} & \textbf{$\SI{59.52}{\percent}$} \\
    GPT2                        & $\SI{88.69}{\percent}$          & $\SI{45.22}{\percent}$          \\
    \bottomrule
  \end{tabular}
  \caption{Comparison of beat ($f_\text{b}$) and downbeat ($f_\text{db}$) F1-scores for different NLP task interpretations. The results demonstrate the superiority of the T5 transformer model, which frames beat tracking as a translation task, over the GPT-2 model, which treats it as a text completion problem.}
  \label{tab:nlp_task}
\end{table}

\subsubsection{Effect of Augmentation}
\label{sec:abl_augmentation}

One of the main advantages of using symbolic MIDI input over audio is that it is fairly easy and efficient to augment the examples dynamically during training. In this experiment, we evaluate the effect of different augmentation methods for pitch and time.
As shown in Table \ref{tab:augmentation_results}, the transposition of the notes' pitches leads to a minor improvement of both downbeat and beat F1-scores.
Applying a randomly drawn time shift alone does not have a noticeable positive effect. Although, in combination with the time scaling augmentation, the best results can be achieved with an $f_\text{b}$ of over $\SI{98}{\percent}$. But this improvement comes at the price of a significantly decreased $f_\text{db}$ score.

\begin{table}[h]
  \centering
  \begin{tabular}{lcc}
    \toprule
    \textbf{Augmentation Methods} & $f_\text{b}$                    & $f_\text{db}$                   \\
    \midrule
    None                          & $\SI{96.03}{\percent}$          & $\SI{59.52}{\percent}$          \\
    Transpose                     & $\SI{96.22}{\percent}$          & \textbf{$\SI{59.91}{\percent}$} \\
    Transpose / Shift             & $\SI{96.06}{\percent}$          & $\SI{58.67}{\percent}$          \\
    Transpose / Shift / Scale     & \textbf{$\SI{98.10}{\percent}$} & $\SI{52.25}{\percent}$          \\
    \bottomrule
  \end{tabular}
  \caption{Impact of different data augmentation strategies on beat ($f_\text{b}$) and downbeat ($f_\text{db}$) F1-scores. The results highlight the effectiveness of pitch transposition (Transpose), time shifting (Shift) and time scaling (Scale), while also showing the tradeoff between improved beat tracking accuracy and decreased downbeat performance.}
  \label{tab:augmentation_results}
\end{table}

\subsubsection{Time Quantization}
\label{sec:abl_quantization}

The temporal resolution has a significant impact on the vocabulary size and offers a tradeoff between precise beat time output and the balancing of the dataset. With a higher temporal resolution, the individual time tokens are less frequently present in the training dataset.
Another effect of quantizing the time tokens is that this way beat annotation inaccuracies get reduced.
Therefore, we highlight the results for different time quantization settings in Table \ref{tab:quantization_results}.
The evaluation shows that there is a noticeable impact on the beat F1-score and a big impact on the downbeat F1-score. By increasing the time steps to $\SI{50}{\milli\second}$, we get the best results for $f_\text{b}$ with $\SI{2}{\percent}$ increase over the result for $\SI{10}{\milli\second}$. By increasing the time steps even to $\SI{100}{\milli\second}$, we see a $\SI{34}{\percent}$ increase in $f_\text{db}$. But these results should be seen with caution since they also benefit from the relatively loose tolerance of $\SI{70}{\milli\second}$ used by mir\_eval. By increasing the quantization step further to $\SI{200}{\milli\second}$, we see a rapid drop of both scores, which matches our expectation since the temporal resolution is too coarse for the evaluation tolerance window now.

\begin{table}[h]
  \centering
  \begin{tabular}{lcc}
    \toprule
    \textbf{Temporal Resolution} & $f_\text{b}$                    & $f_\text{db}$                   \\
    \midrule
    $\SI{5}{\milli\second}$      & $\SI{94.52}{\percent}$          & $\SI{54.36}{\percent}$          \\
    $\SI{10}{\milli\second}$     & $\SI{96.03}{\percent}$          & $\SI{59.52}{\percent}$          \\
    $\SI{20}{\milli\second}$     & $\SI{96.98}{\percent}$          & $\SI{65.17}{\percent}$          \\
    $\SI{50}{\milli\second}$     & \textbf{$\SI{97.88}{\percent}$} & $\SI{73.58}{\percent}$          \\
    $\SI{100}{\milli\second}$    & $\SI{97.68}{\percent}$          & \textbf{$\SI{80.00}{\percent}$} \\
    $\SI{200}{\milli\second}$    & $\SI{71.85}{\percent}$          & $\SI{55.35}{\percent}$          \\
    \bottomrule
  \end{tabular}
  \caption{Effect of different time quantization settings on beat ($f_\text{b}$) and downbeat ($f_\text{db}$) F1-scores. The results demonstrate how increasing the quantization step can enhance downbeat detection but also highlight the performance drop when the resolution becomes too coarse.}
  \label{tab:quantization_results}
\end{table}

\subsection{Comparison with Baselines}
\label{sec:comparison}

\begin{table*}[t]
  \centering
  \begin{tabular}{l|cc|cc|cc|cc}
    \toprule
    \textbf{Method}                             & \multicolumn{2}{c|}{\textbf{A-MAPS}} & \multicolumn{2}{c|}{\textbf{ASAP}} & \multicolumn{2}{c|}{\textbf{GuitarSet}} & \multicolumn{2}{c}{\textbf{Leduc}}                                                                                                                                         \\
    \cmidrule{2-9}
                                                & $f_\text{b}$                         & $f_\text{db}$                      & $f_\text{b}$                            & $f_\text{db}$                      & $f_\text{b}$                    & $f_\text{db}$                   & $f_\text{b}$                    & $f_\text{db}$                   \\
    \midrule
    Beat This! \cite{foscarin2024beatthis}      & -                                    & -                                  & $\SI{76.30}{\percent}$                  & $\SI{61.20}{\percent}$             & $\SI{92.20}{\percent}$          & $\SI{88.10}{\percent}$          & -                               & -                               \\
    \midrule
    HMM (J-Pop) \cite{shibata2021non-local}     & $\SI{48.63}{\percent}$               & $\SI{25.62}{\percent}$             & $\SI{47.68}{\percent}$                  & $\SI{13.36}{\percent}$             & $\SI{38.37}{\percent}$          & $\SI{6.78}{\percent}$           & $\SI{32.71}{\percent}$          & $\SI{13.38}{\percent}$          \\
    HMM (classical) \cite{shibata2021non-local} & $\SI{49.85}{\percent}$               & $\SI{28.23}{\percent}$             & $\SI{43.60}{\percent}$                  & $\SI{13.67}{\percent}$             & $\SI{33.65}{\percent}$          & $\SI{9.88}{\percent}$           & $\SI{34.34}{\percent}$          & $\SI{13.78}{\percent}$          \\
    PM2S \cite{liu2022performance}              & $\SI{83.89}{\percent}$               & $\SI{68.90}{\percent}$             & \textbf{$\SI{82.95}{\percent}$}         & $\SI{14.14}{\percent}$             & $\SI{42.63}{\percent}$          & $\SI{16.49}{\percent}$          & $\SI{41.12}{\percent}$          & $\SI{28.94}{\percent}$          \\
    \midrule
    \textbf{Ours}                               & \textbf{$\SI{98.01}{\percent}$}      & \textbf{$\SI{76.56}{\percent}$}    & $\SI{78.13}{\percent}$                  & \textbf{$\SI{27.81}{\percent}$}    & \textbf{$\SI{52.38}{\percent}$} & \textbf{$\SI{23.02}{\percent}$} & \textbf{$\SI{57.72}{\percent}$} & \textbf{$\SI{29.75}{\percent}$} \\
    \bottomrule
  \end{tabular}
  \caption{Comparison of beat ($f_\text{b}$) and downbeat ($f_\text{db}$) F1-scores between the proposed transformer-based model and state-of-the-art MIDI beat tracking methods, including an HMM-based approach \cite{shibata2021non-local} and the PM2S deep learning model \cite{liu2022performance}. For reference, results from the audio-based Beat This! model are also included. The evaluation is conducted on the test splits of the A-MAPS, ASAP, GuitarSet, and Leduc datasets, highlighting differences in performance across piano and guitar music.}
  \label{tab:sota_comparison}
\end{table*}

Using the results of the ablation study, we optimized our final model for a comparison with other state-of-the-art methods. We choose a segment length of $\SI{10}{\second}$, a temporal resolution of $\SI{50}{\milli\second}$ and we apply all three augmentation methods. The model is trained on a combined training dataset containing the respective train splits of the A-MAPS, ASAP, GuitarSet, and Leduc datasets. Since velocity is not included in the guitar datasets, the encoding \textit{v2} is used.

We compare our model's performance with two state-of-the-art MIDI beat tracking approaches: The strongest HMM-based approach \cite{shibata2021non-local} and the PM2S deep learning model \cite{liu2022performance} which relies on neural beat tracking. For reference, we also add the evaluation results of Beat This!, the currently best-performing audio-based beat tracking transformer model. The scores for Beat This! are directly taken from the 8-fold cross-validation results of the original paper and are therefore intended to give an impression rather than a fair comparison.
The results of the comparative evaluation on the test splits of the four datasets are shown in Table \ref{tab:sota_comparison}.

Our proposed transformer-based model outperforms PM2S and the HMM on almost every dataset for beat as well as downbeat performance. Only on the ASAP dataset, the beat tracking results of PM2S are $\SI{6}{\percent}$ better compared to our results.
In general, we can see a significant drop in downbeat accuracy when comparing the results for the A-MAPS dataset with the others. This is an indicator that the data is easier to learn, which is caused by the way the dataset has been generated (see Section \ref{sec:discussion}). We also see a significant drop in $f_\text{b}$ when comparing guitar with piano dataset results. This indicates that detecting beats in guitar music is more difficult, since we have generally a lower note density and less strong downbeat indication, especially in guitar solos.
Since the pieces in the ASAP dataset consist of more complex time-signatures than in the A-MAPS dataset, we see a much higher drop in the downbeat than in the beat F1-score.

The audio-based method performs generally better in terms of downbeat F1-score, since a lot of expression gets lost when transcribed as a MIDI file. The accents that typically are placed on the downbeat make it a lot easier to determine the beats' positions. The huge differences in the GuitarSet dataset are most likely caused by the inaccuracies of the MIDI annotations and the discrepancy with the beat annotations (see Section \ref{sec:discussion}).
These might also be responsible for the higher F1-scores for the Leduc guitar transcription dataset. Although the Leduc dataset consists of Jazz pieces with more complex rhythms, the alignment between MIDI and beat annotations is better because of the joint beat and note alignment process described in Section \ref{sec:datasets}.

\subsection{Discussion}
\label{sec:discussion}

While there are datasets available containing performance MIDI with beat annotations, the quality strongly varies. Since the A-MAPS MIDI files are derived from tempo-varied quantized MIDI files, they consequently do not capture the full range of human timing variations \cite{foscarin2020asap}. On the other hand, they offer a perfect alignment between beat and note annotations.
The GuitarSet examples consist of actual guitar recordings that have been transcribed semi-automatically using a hexaphonic pickup and manual corrections \cite{xi2018guitarset}. While the note annotations have been carefully adjusted, the beat annotations come from the used metronome and do not account for any tempo variations by the human player.
Datasets like Leduc and ASAP have an improved alignment workflow and lead to a better agreement between beat and note annotations.

\section{Conclusion}\label{sec:conclusion}
This research demonstrates the effectiveness of an end-to-end transformer-based approach for beat tracking in MIDI performances. By formulating the task as a symbolic translation problem, our model surpasses existing MIDI-based methods, including HMM-based approaches and PM2S, achieving state-of-the-art performance across most datasets. Key contributions include a data pre-processing pipeline, data augmentations, and tokenization strategies.
The model's adaptability to diverse datasets for guitar and piano highlights its generalizability. Augmentation strategies and temporal quantization enhance the beat tracking accuracy. Although its accuracy does not yet match that of audio-based beat tracking methods, the model remains highly effective in scenarios where only MIDI data is available, offering a practical alternative for rhythm analysis in symbolic music.

In future works, this method could be combined with a AMT approach in order to have a more fair comparison with audio based beat-tracking methods. By combining this method with a beat based quantization method, it could be used in a processing pipeline that transcribes audio to sheet music. Expanding the model to multi-instrument performances and applying self-supervised learning techniques could also increase its robustness and versatility in diverse musical settings.


\bibliography{smc2025_midibeats}

\end{document}